\documentclass[prb,twocolumn,superscriptaddress,showpacs,amsmath,amssymb]{revtex4-1}

\usepackage{graphicx}
\usepackage{latexsym}
\usepackage{amsmath}
\usepackage{amssymb}
\usepackage{amsfonts}
\usepackage{color}
\usepackage{bm}
\usepackage{verbatim}
\usepackage{ulem}

\DeclareMathAlphabet\mathbfcal{OMS}{cmsy}{b}{n}

\definecolor{MS-color}{RGB}{128,0,128}

\usepackage{framed}
\definecolor{shadecolor}{RGB}{222,222,221}
\begin{document}

\title{Chiral pair density wave states generated by   spin supercurrents }

\date{\today}

\author{M.A.~Silaev}
 \affiliation{Department of
Physics and Nanoscience Center, University of Jyv\"askyl\"a, P.O.
Box 35 (YFL), FI-40014 University of Jyv\"askyl\"a, Finland}

\affiliation{Moscow Institute of Physics and Technology, Dolgoprudny, 141700 Russia}

\author{D. S. Rabinovich}
\affiliation{Skolkovo Institute of Science and Technology, 3 Nobel Street, Moscow, Russia 121205}

\author{I. V. Bobkova}
\affiliation{Institute of Solid State Physics, Chernogolovka, Moscow
  reg., 142432 Russia}
\affiliation{Moscow Institute of Physics and Technology, Dolgoprudny, 141700 Russia}
\affiliation{National Research University Higher School of Economics, Moscow, 101000 Russia}

 \begin{abstract}
 We report that spin supercurrents in magnetic superconductors and superconductor/ferromagnetic insulator bilayers can induce the  Dzyaloshinskii–Moriya interaction  which strength is proportional to the superconducting order parameter amplitude.  This effect leads to the existence of inhomogeneous parity-breaking ground states  combining the chiral magnetic helix and the pair density wave orders. The formation of such states takes place via the penetration of chiral domain walls at the threshold temperature
 below the superconducting transition.
  We find regimes with both the single and the re-entrant transitions into the inhomogeneous states with decreasing temperature.  
 The predicted hybrid chiral  states can be found in the existing structures with realistic parameters and materials combinations.   
 \end{abstract}

 \maketitle

   Recently the interest in    superconductor/ferromagnetic insulator (S/FI) structures has been growing rapidly.
 Experimental advances in fabrication of S/FI systems 
 has resulted in the observation of the spin-split spectrum of Bogolubov quasiparticles  in  Al/EuS systems \cite{Hao1990,Li2013, strambini2017revealing,
PhysRevResearch.3.023131}. The  combination of the effective Zeeman field generated by the magnetic proximity effect and superconductivity leads to the long-range spin transport in EuS/Al systems\cite{Wolf2014,Bobkova2015,Silaev2015,Bobkova2016}, non-trivial spin pumping effects in  NbN/GdN \cite{yao2018probe} and Nb/yttrium iron garnet (YIG)  \cite{Jeon2020giant} structures.
Moreover recently the InAs/Al/EuS multilayers combining  superconductivity, ferromagnetism and spin-orbit coupling (SOC) have been proposed as the platforms for the realization of the  topological superconductivity\cite{vaitiekenas2021zero,
manna2020signature}. 

\begin{figure}[htb!]
  \centerline{$
 \begin{array}{c}
 \includegraphics[width=2.8in]
 {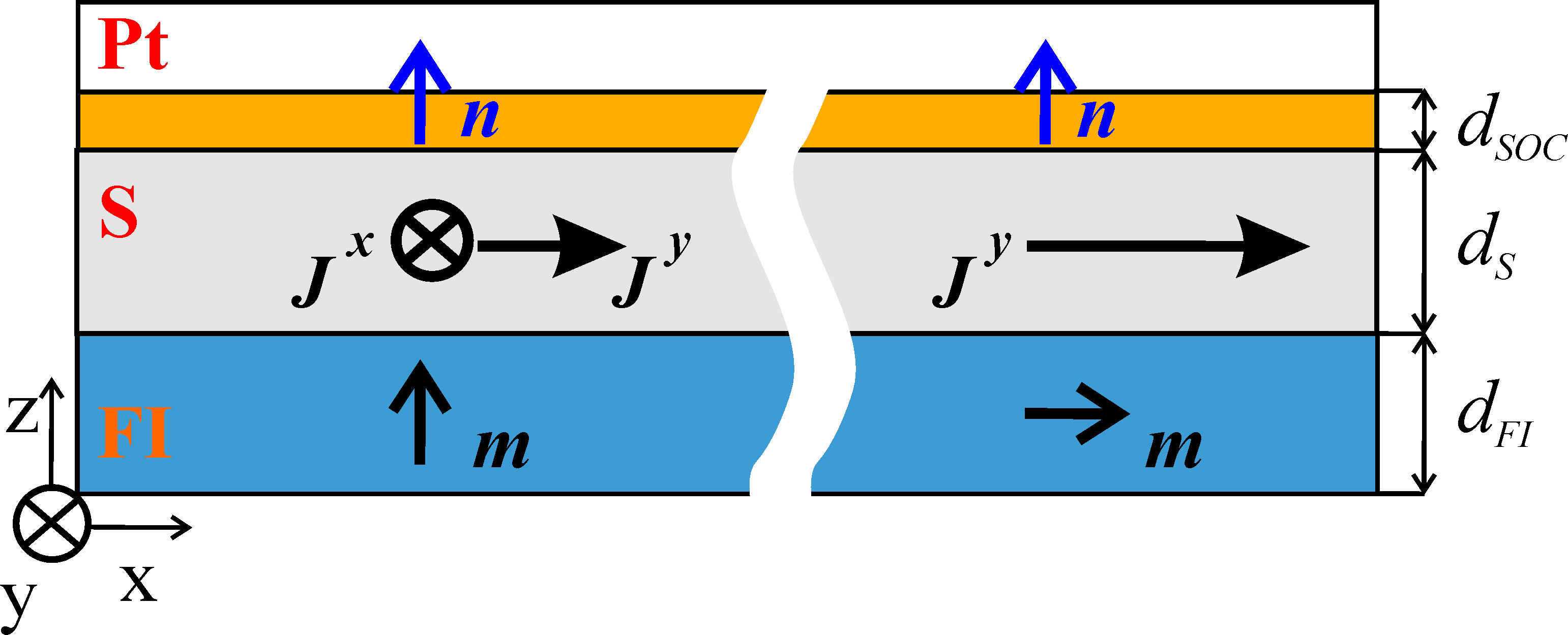}
  \end{array}$}
 \caption{\label{Fig:Setup}
 Geometry of the multilayered  material consisting of S film in contact with heavy metal Pt layer having interfacial Rashba-type SOC vector $\bm n$ and FI inducing the Zeeman field in S. There is a superconducting spin current $J_j^\gamma$ in S which magnitude and direction depends of the direction of the magnetic moment $\bm m$. The spin torque induced by the gradients of $J_j^\gamma$ leads to the DMI (see text).   }
  \end{figure}

In this Letter we demonstrate that the interplay of  superconductivity, ferromagnetism and SOC generates a non-trivial ground state characterized by the chiral spin order accompanied by the spatial modulation of the superconducting order parameter. 
Such state which we call the chiral pair density wave (CPDW) state can occur in S/FI systems with interfacial SOC induced by heavy metal Pt layers as shown in Fig.\ref{Fig:Setup} as well as in a wider class of magnetic superconductors with spin-singlet order parameter\cite{
stolyarov2018unique,
devizorova2019superconductivity,
iida2019coexisting,
smylie2018anisotropic} .

The CPDW states {break the parity symmetry and } have several  other qualitatively different features compared to the previously studied cryptoferromagnetic states \cite{anderson1959spin} in the weakly magnetic superconductors\cite{bulaevskii1985coexistence} and ultra-thin metallic ferromagnetic films on top of the superconductor\cite{ buzdin1988ferromagnetic,bergeret2000nonhomogeneous}. 

The mechanism which provides the formation of CPDW state is the  Dzyaloshinskii-Moriya interaction\cite{dzyaloshinskii1964theory,
dzyaloshinsky1958thermodynamic,
PhysRev.120.91} (DMI) generated by the superconducting spin currents.
The corresponding  DMI interaction energy  has the form 
 \begin{align} \label{Eq:DMGen}
  F_{DM} =  |\psi|^2 \bm d_k  \cdot (\bm m \times \nabla_k \bm m)
 \end{align}
 where  $\bm m= \bm m (\bm r)$ is the magnetic texture, 
 $\bm d_{1,2,3}$ are the DMI vectors and $\psi$ is the s-wave spin-singlet superconducting order parameter. 
The  interaction between the superconductivity and magnetism provided by  Eq.(\ref{Eq:DMGen}) is
determined by the magnetization gradients. Therefore it is conceptually distinct from the mechanism which involves superconducting phase gradients discussed in the context of the Edelstien effect\cite{edelshtein1995, edelstein2005,PhysRevB.67.020505,edelstein2021ginzburg}, helical superconducting states\cite{kaur2005helical,
samokhin2005paramagnetic}, spontaneous currents\cite{Bobkova2004,Mironov2017} and the anomalous Josephson effect\cite{Buzdin2008}. 


The DMI provides  the formation of magnetic spirals and skyrmions  \cite{bogdanov1989thermodynamically,bogdanov1994thermodynamically} which are considered as the building blocs of the future spintronics devices \cite{Fert2013}. Therefore controlling DMI with external parameters such as the voltage and strain attracts significant attention \cite{
sadovnikov2018magnon,gusev2020manipulation,
shibata2015large,
liu2017chopping}. 
The superconductivity-induced DMI (\ref{Eq:DMGen}) provides a  plethora of different effects resulting in the control over the magnetic textures with the help of the superconducting correlations. 

{ To start with let's demonstrate that the  DMI interaction of the form (\ref{Eq:DMGen}) is induced by spin supercurrents.}
 Consider a spin-singlet superconductor  with  the Zeeman field $\bm h (\bm r) = h  \bm m(r)$, where $|\bm m| =1$ and the linear in momentum  SOC potential $V_{so}=  \alpha_{SO} {\cal A}_{k}^\gamma \hat\sigma_\gamma p_k $. Here $\hat{\bm \sigma}$ and $\bm p$ are the electron spin and momentum, $ \alpha_{SO} $ is the SOC strength and ${\cal A}_{k}^\gamma \hat\sigma_\gamma$ is the global SU(2) field.
 E.g. for the Rashba-type SOC ${\cal A}_{k}^\gamma = n_j \varepsilon_{j\gamma k}$, where $\bm n$ is the anisotropy vector.

 Our first main result is that the Ginzburg-Landau free energy of this system can be written as follows\cite{supplement} 
 \begin{align} 
  \label{Eq:EnergyGradientSGL}
 &  F_S  =  a \psi^2 + 
 \frac{b}{2} \psi^4  + c (\nabla \psi) ^2
 -
 \frac{a_m}{2} ({\cal{D}}_k \bm m)^2 \psi^2 
 \end{align}
 %
 Coefficients $a$, $b$, $c$ determine the usual GL free energy in a diffusive superconductor with the Zeeman splitting. 
 
 The last term in (\ref{Eq:EnergyGradientSGL}) provides the coupling between superconductivity and ferromagnetism determined by the covariant derivative of the magnetization
 \cite{Tokatly2008,10.21468/SciPostPhys.10.3.078}  ${\cal D}_k \bm m = ( \nabla_k - \alpha_{SO} {\cal{\bm A}}_k \times ) \bm m$, where 
 $ {\mathbfcal{A}_k} = ({\cal A}_{k}^x, {\cal A}_{k}^y, {\cal A}_{k}^z) $.
 The coefficient is  
 $ a_m= \pi  \nu D  T_c^3 h^2 \sum_{\omega>0} 
 ( h^2 + \omega^2)^{-2}  $, where $\omega$ are the Matsubara frequencies. By the order of magnitude $a_m \sim  \nu  T_c^2 \xi^2_0 (h/T_c)^2$, { where $T_c$ is the superconducting critical temperature and $\nu$ is the normal state density of states in the superconductor.} 
  The physical meaning of this term is the
 twisted exchange interaction between spins  mediated by the Cooper pairs in S. It generalizes seminal de Gennes result\cite{de1966coupling} on the renormalization of RKKY interactions in superconductors to the case when 
 the SOC is present. 
{Quite importantly, $a_m>0$ so that the gradients of $\bm m$ become less energetically costly. In the absence of SOC this leads eventually to the negative spin stiffness resulting in the cryptoferromagnetic state\cite{anderson1959spin,bergeret2000nonhomogeneous}. In the presence of SOC the situation is more complicated because as we show below the CPDW state appears even for the overall positive  spin stiffness.}

 The last term in Eq.~(\ref{Eq:EnergyGradientSGL}) can be interpreted as the energy of spin supercurrent. 
 Indeed using the  Usadel equation and the recently obtained general quasiclassical expression for the superconducting free energy \cite{virtanen2020quasiclassical} we obtain the general expression for the equilibrium spin current in the superconductor with spin-singlet pairing  \cite{supplement}
 \begin{align}
 \label{Eq:SpinCurrentVar}
 J^\gamma_k = \frac{\delta}{\delta {\cal A}_k^\gamma} \int d\bm r F_{S}
 \end{align}
 The equilibrium spin
 supercurrent Eq.~(\ref{Eq:SpinCurrentVar}) provides a
 field-like spin torque acting on the magnetic texture $\bm m (\bm r)$. The correction to the effective field 
from superconducting correlations is $\delta \bm H_{eff} = - \delta F_{S}/\delta \bm M $. The corresponding spin torque is given by  $\bm T = \gamma \bm m\times \delta \bm H_{eff}$
and can be written as \cite{tokatly2019universal}
     \begin{align} \label{Eq:T}
    & \bm T  = 
    \gamma {\cal D}_k\bm J_k 
    \end{align}
    where $\bm J_k = (J_k^x, J_k^y, J_k^z)$. 
    This expression agrees with the general symmetry of a gauged Heisenberg model under the local spin rotations \cite{Tokatly2008,10.21468/SciPostPhys.10.3.078}.
   
  The covariant magnetic energy in Eq.~(\ref{Eq:EnergyGradientSGL})
  has three qualitatively different contributions
  $({\cal D}_k\bm m)^2 = (\nabla_k\bm m)^2  + \alpha_{SO}^2({\mathbfcal{A}_k} \times \bm m)^2  - 
  2 \alpha_{SO}{\mathbfcal{A}_k 
  (\bm m \times \nabla_k \bm m )}
  $. The first two terms provide renormalization of the usual exchange\cite{de1966coupling, bergeret2000nonhomogeneous,  aikebaier2019superconductivity} and anisotropy energies which are present already in the free energy of the ferromagnet $ F_{FI} = 
  [A (\nabla \bm m)^2 + K m_z^2]/2$.
  Here we assume the { $K$ can have arbitrary sign, so that in the normal state the structure is in-plane (out-of plane) magnetized $\bm m\parallel \bm x$ ($\bm m\parallel \bm z$). As we show below the superconducting spin current energy in (\ref{Eq:EnergyGradientSGL})  typically yields the easy-axis anisotropy correction which can overcome the intrinsic easy-plane one in case is $K>0$ and lead to the magnetic reorientation transition.  }
  The change of anisotropy in the superconducting state of S/FI systems has been recently observed experimentally\cite{gonzalez2021superconductivity,gonzalez2020superconductivity,PhysRevApplied.14.024086,tikhomirov2021anisotropy}. 
 The third term is the DMI in the form (\ref{Eq:DMGen}) with vectors
 \begin{align} \label{Eq:DMIvector}
 \bm d_k =2 a_m \alpha_{SO}  {\mathbfcal{A}}_k .
 \end{align}
 This DMI appears only in the superconducting state with SOC because of the non-local twisted spin response mediated by the condensate. The non-local character of the DMI in the considered system provides the {\it chiral coupling} between the order parameter and magnetization which has not been taken into account in the previous works on the ferromagnetic superconductivity.

 In order to feature such an interaction the system should have three basic ingredients:
 usual spin-singlet superconducting pairing, exchange field and SOC. 
 The generic device  of such type  is shown in Fig. \ref{Fig:Setup}.
 It consists of the usual S layer sandwiched between the FI and the heavy
 metal Pt films. The S thickness $d_S$ is much larger than the atomic scale but smaller than the superconducting coherence length
 $\xi_0 = \sqrt{D/T_c}$, where $D$ is the diffusion coefficient in S. 

  The exchange field in  S is induced in result of the spin-mixing scattering of electrons at S/FI interface \cite{Tokuyasu1988,
cottet2009spin,
Moodera1988,
Hao1990,
manna2020signature,
Li2013,
strambini2017revealing,
PhysRevResearch.3.023131,
vaitiekenas2021zero}. For thin S film $d_S< \xi_0$, where $\xi_0$ is the zero-temperature coherence length the effective Zeeman field is
 $\bm h = (J_{sd}/d_S) \bm m$. Here $J_{sd}$ is the interfacial exchange interaction constant\cite{ohnuma2014enhanced} proportional to the spin-mixing angle\cite{Tokuyasu1988,
cottet2009spin}. 

 The SOC is generated at the S/Pt interface within the layer which thickness is smaller than that of the superconducting film $d_{SOC}< d_S$. Therefore we will use the interfacial SOC potential
 $\tilde \alpha_{SO} = \alpha_{SO} d_{SOC} $.
 Further we  focus on a particular case of {\it Rashba SOC}
 when the only non-zero components of the SU(2) field are 
 $ {\cal A}_{y}^x = - {\cal A}_{x}^y = 1 $. Then the free energy (\ref{Eq:EnergyGradientSGL}) yields the DMI in the form
(\ref{Eq:DMGen}) with the vectors $\bm d_k = (d_k^{(1)}, d_k^{(2)}, d_k^{(3)})$
 \begin{align} \label{Eq:DMIRashba}
  & d_k^{(\gamma)} =   
  \frac{2\tilde \alpha_{SO} a_m}{d_{FI}}  \varepsilon_{z\gamma k}  
   \end{align}
   The correction to the anisotropy energy
has an easy-axis form  and renormalizes the anisotropy constant to { $\tilde K =K - \tilde{\alpha}_{SO}^2 a_m \psi^2 d_{FI}d_{SOC}$.}    
   Besides that Eq.(\ref{Eq:EnergyGradientSGL}) yields the change  of the exchange stiffness 
 $\tilde A = A- a_m \psi^2 d_S /d_{FI}$.

 For a thin FI film superconducting corrections 
 can renormalize the exchange stiffness and anisotropy rather strongly.
 The general reason is that the condensation $\nu T_c^2 \sim 10^6$ erg/cm$^3$
 is much larger than the typical magnetic anisotropy energy 
  $|K| \sim 10^3$ erg/cm$^3$. 
 Here we assume that  the Fermi wavevector of the superconductor $k_F = 10^8$ cm$^{-1}$, 
  $T_c=10$ K, $E_F=10^4$ K. 
 Then we get the 
 $a_m \sim \xi_0^2 (h/T_c)^2 \nu T_c^2 $ or 
 $a_m \sim (T_c/\xi_0) (h/T_c)^2 (k_F\xi_0)^3 (T_c/E_F) $
 which yields $a_m\sim 10^{-7} (h/T_c)^2$ erg/cm. 
  Provided that $h\sim 0.5 T_c$ this correction is of the same order as the characteristic value of the stiffness constant in YIG \cite{klingler2014measurements,
stancil2009spin} $A\sim  10^{-7}$ erg/cm. The effective value of magnetic stiffness $  (d_{FI}/d_S)A$  can be reduced by reducing the relative thickness of FI. 
This large correction provides significant change of the domain wall size\cite{aikebaier2019superconductivity} in FI below $T_c$. In combination with the effect of DMI it eventually allows for the chiral ground state in the setup shown in Fig.\ref{Fig:Setup}.

 \begin{figure}[htb!]
   \centerline{$
 \begin{array}{c}
 \includegraphics[width=2.8in]
  {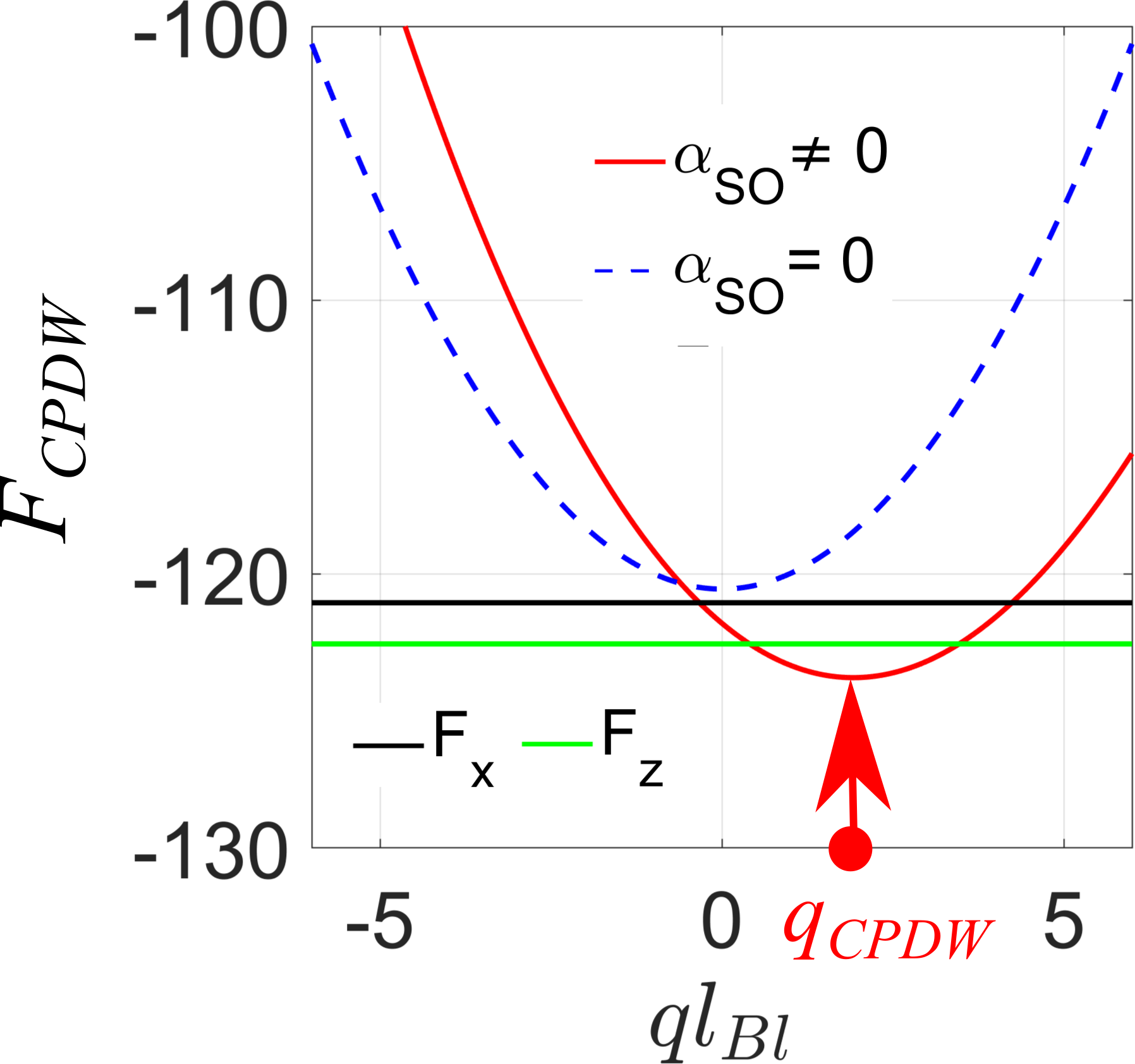}
  \end{array}$}
 \caption{\label{Fig:Energy}
  (Red solid line): The free energy of the CPDW probe state with the uniformly rotated magnetization in the presence of SOC $\alpha_{SO}\neq 0$.  The equilibrium wave number is $q_{CPDW}$. 
  (Blue dashed line): The same system in the absence of SOC. 
  (Black, green lines) : Energy levels $F_x$ and $F_z$ of the homogeneous states  with $\bm m \parallel \bm x$ and $\bm m \parallel \bm z$, respectively. The temperature is $T=0.75 T_c$ and the other parameters are defined in the text. The plot corresponds to easy-axis anisotropy $K<0$. The case although $K>0$ looks very similar. 
  }
  \end{figure}

  %
 Let us estimate the change of the domain wall energy induced by DMI.
Taking into account that the anisotropy energy is 
$K=4\cdot 10^3$ erg/cm$^3$ we get the typical Bloch domain wall thickness
$l_{Bl} = \sqrt {A/K} \sim 10^{-5} $
 cm and energy per unit length in the absence of DMI 
 $E_{Bl} \sim K d_{FI} l_{Bl}\sim 10^{-8}$ erg/cm.  
  Taking $\tilde \alpha_{SO} \sim 0.1-1$ 
 \cite{lo2014spin,banerjee2018controlling, ast2007giant}    we get the estimation of the DMI  
   $F_{DM}\sim |\psi|^2 a_m/(d_{FI} l_{Bl})$ which yields $F_{DM}\sim (0.1/ l_{Bl})  (h/T_c)^2 $ erg/cm$^3$ provided that $|\psi| \sim 1$. 
 The DMI contribution to the domain wall energy is $\delta E_{Bl} \sim 0.1 d_{FI} (h/T_c)^2 \sim 10^{-7} (h/T_c)^2$ erg/cm.  For $h/T_c \sim 1$ it is of the same order or larger than the domain wall energy without the DMI. 
 Thus,  superconductivity-controlled DMI \ref{Eq:DMGen}   is capable of making the  domain wall energy negative leading to the  formation of non-homogeneous magnetic ground states\cite{bogdanov1994thermodynamically}. The back-action of the magnetic texture leads to the the spatial modulation of the  order parameter amplitude -  the pair density wave state. 

   To study the  hybrid ground state combining the {  pair density wave and  the magnetic helix} we minimize numerically the energy given by $F= d_SF_S + d_{FI}F_{FI}$.
  Introducing the spherical angles for magnetization direction 
  $m_x = \sin\theta \cos \chi$, 
$m_y = \sin\theta \sin \chi$, $m_z = \cos\theta $, we write the magnetic $F_{FI}$ and superconducting $F_S$ energies setting $\chi=0$  and introducing normalized units $x/l_{Bl}$, 
 $\tilde F_{FI} = F_{FI} /|K| $,
 $\tilde F_{S} = F_{S} / |K| $
 \begin{align} \label{Eq:EnergyFI}
 & \tilde F_{FI} = 
 (\partial_x\theta)^2  
  \pm   \cos^2\theta  
  \\ \label{Eq:EnergyS}
 &  \tilde F_{S}  =  
 \left(\frac{l_{Bl}}{\xi_0}\right)^2 
 [ \tilde a \psi^2 + 
 \frac{\tilde b}{2} \psi^4 ] + 
  \tilde c  
 (\partial_x \psi) ^2  - 
  \frac{\tilde a_m}{2} \psi^2 \times
 \\ \nonumber
  &  \left[ (\partial_x\theta)^2 
+  \left(
  \frac{2\tilde\alpha_{SO}  l_{Bl}}{d_S} \right )
 \partial_x \theta 
 +
 \left(\frac{\tilde\alpha_{SO}  l_{Bl}}{d_S}\right)^2
 \frac{d_S}{d_{SOC}} \cos^2\theta\right]  
   \end{align}
   where upper (lower) sign corresponds to the easy plane (axis)
   anisotropy $K>(<) 0$.
   At $h=0.5 T_c$ the numerical values are 
 $\tilde a = \beta( T/T_c -1) $, 
$\tilde b =  0.046 \beta $, 
$\tilde c = 0.37 \beta $, $\tilde a_m = 0.008 \beta$.  where $\beta = \nu T_c^2\xi_0^2 /A \sim 10$.
   Here we assume that FI film is thin and neglect the effect of stray fields both on the superconductivity and on the magnetic energy. This provides much more preferable conditions for the nucleation of the chiral phase than in the bulk chiral magnet \cite{bogdanov1994thermodynamically}. 
 
 {To demonstrate the possibility of the chiral state formation we consider the probe function with the homogeneously rotating magnetization and $\psi= const$. 
  For simplicity we put $d_{FI}=d_S=d_{SOC}=\xi_0$ in Eq.(\ref{Eq:EnergyFI},\ref{Eq:EnergyS}).
  After minimizing the free energy by $\psi$ we get $F_{CPDW}(q)=q^2 \pm 1/2 - (\tilde a-a_m(\tilde\alpha^2_{SO}/4 + q^2/2+\tilde\alpha_{SO} q))^2/2\tilde b$. It has to be compared with the two possible homogeneous configurations with $F_x= -\tilde a^2/2\tilde b$ for $\bm m =\pm  \bm x$ and $F_z =\pm 1 - (\tilde a -\tilde a_m \tilde\alpha_{SO}^2/2)^2/2\tilde b $
 for $\bm m = \pm \bm z$. The CPDW state with necessity occurs if $F_{CPDW}(q=0)< F_{x,z}$
 which is realized under the condition $3\tilde\alpha^2_{SO} a_m /8 - 3 \tilde b /\tilde \alpha^2_{SO} a_m<a< \tilde\alpha^2_{SO} \tilde a_m /8 - 2 \tilde b /\tilde\alpha^2_{SO} a_m$. 
 Note that this condition  can be realized
for the positive overall spin stiffness $\partial^2 F/\partial^2 q >0$ in contrast to cryptofferomagnetic state.  In case if either $F_{CPDW}(q=0)>F_{x}$ or $F_{CPDW}(q=0)>F_{z}$ the homogeneous state can be more preferable. 
This condition is sufficient but not necessary for the realization of the CPDW. 
The energy profile $F_{CPDW}(q)$ is shown in  Fig.\ref{Fig:Energy} for the typical parameters of the system 
 $\tilde \alpha_{SO} =0.8$, $l_{Bl}=\xi_0$, $h=0.5 T_c$, $\beta=10$.  One can see that although $F_x>F_{CPDW}(q=0) > F_z$ the minimum is given by the CPDW state since $F_{CPDW}(q=q_{CPDW}) < F_{z,z}$.  
 }

{The probe state discussed above is not exact and overestimates the energy because it neglects the variation of $\psi$ and considers only single $q$.} 
To characterise the  CPDW state  we numerically minimize the energy $F$ with boundary conditions $\psi(0)=\psi(L_{pdw})$ and $\theta(0,L_{pdw})=0,\pi$, where the length $L_{pdw}$ is the half-period of the chiral state. 
Note that here we assume $\alpha_{SO}>0$
which determines the positive chirality of the domain walls $[\theta(L_{pdw})-\theta(0)]/\pi =1$. For $\alpha_{SO}<0$ the chirality will be $-1$.
 Then we compare energies of the uniform states
 $F_{x,z}$   vs $F_{CPDW} =  \int_0^{L_{pdw}} dx F /L_{pdw}$. 
For the parameters where  the ground state is the CPDW the energy difference $E_{dw}-E_{un}$ becomes negative. For calculation we choose the easy-plane $K>0$ and the parameters $h=0.5 T_c$,  $\beta=10$.%

\begin{figure}[htb!]
   \centerline{$
 \begin{array}{c}
 \includegraphics[width=1.0 \linewidth]
 {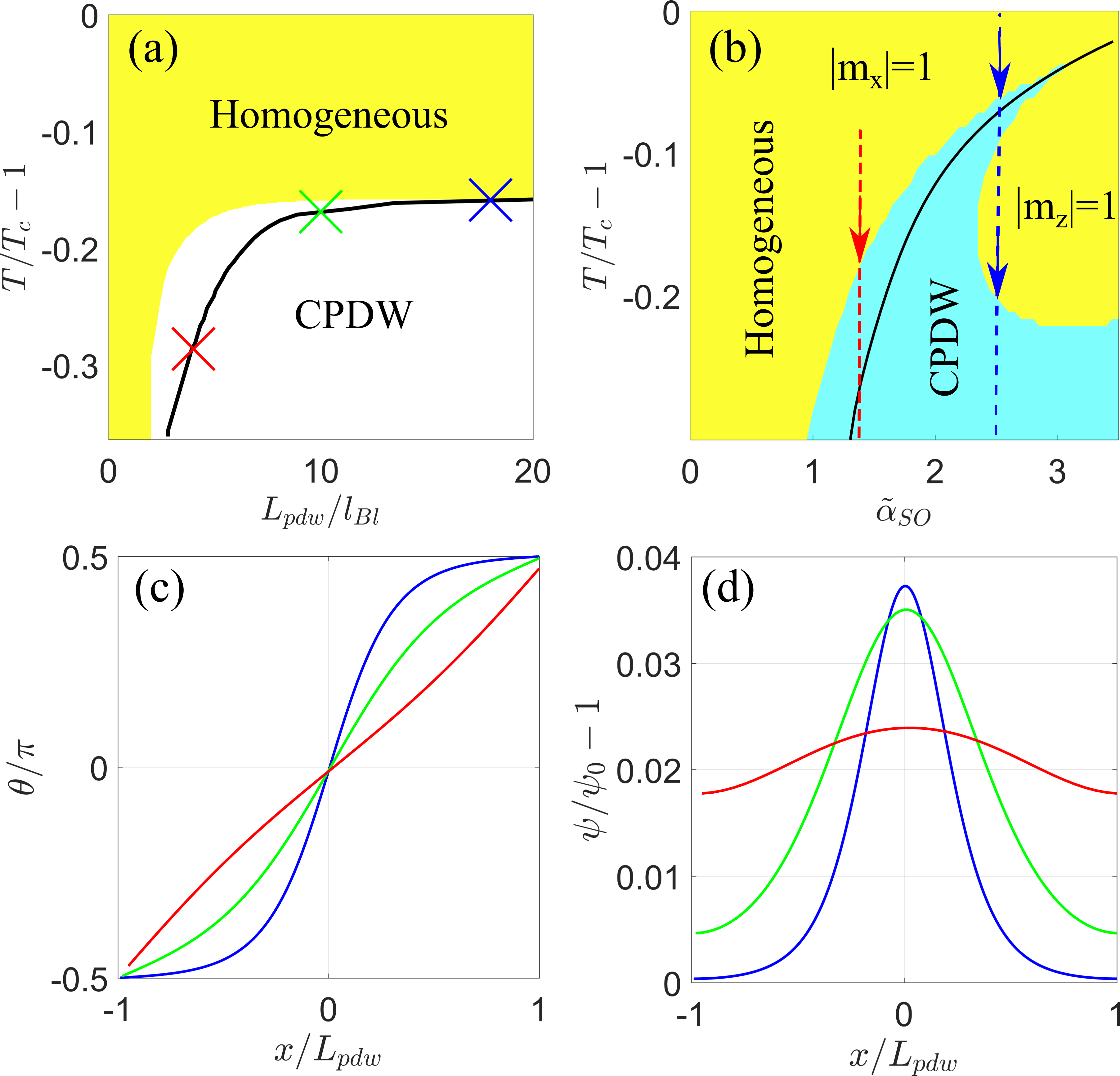}
  \end{array}$}
  \caption{\label{Fig:PDWphase}
  (a) Phase diagram of Pt/S/FI system as function of $(T,L_{pdw})$ with $\bar{\alpha}_{SO} = 1.5$. The black curve shows equilibrium CPDW state period $L_{pdw (T)}$. 
 (b) Phase diagram of Pt/S/FI system as a function of  $(T,\alpha_{SO})$ showing the regions of single transition (e.g. along red dashed line) and the re-entrant behaviour of the CPDW state (e.g. along blue dashed lines). The solid black line shows the points of vanishing effective anisotropy where $F_x=F_z$. 
     (c,d) Structure of PDW state: magnetic part $\theta (x)$ and superconducting part $\psi (x)$. The red, green and blue lines correspond to the parameters marked by the red, green and blue crosses in the panel (a).
  In all panels $d_{FI}=d_S=d_{SOC}=\xi_0$, $l_{Bl} =1.5\xi_0$, other parameters are described in the text. }
  \end{figure}

In Fig.\ref{Fig:PDWphase}a we demonstrate the characteristic phase diagram of the Pt/S/FI system in the plane of $(L_{pdw}, T)$.  The black line shows the equilibrium chiral state half- period $L_{pdw} (T)$. 
One can see that the domain walls first occur at $T_{pdw}$ with the zero concentration, i.e. the infinite inter-wall distance $L_{pdw}(T_{pdw}) = \infty$. 
At lower temperatures  
$T<T_{pdw}$ 
the finite concentration of domain walls, determined by the distance between the neighbour domain walls $L_{pdw}$, appears. In the direct analogy with Abrikosov vortices in type-II superconductor the concentration of domain walls is determined by the  balance between the negative energy of the individual wall and the repulsion between neighbour walls.  %
This can be seen in Fig.\ref{Fig:PDWphase}c,d which show with the red, green and blue curves the evolution of the CPDW state components  $\theta(x)$, $\psi(x)$   with the temperature approaching $T_{pdw}$ as marked by crosses on the black curve $L_{pdw}(T)$ in the panel Fig.\ref{Fig:PDWphase}a.  
The scales of $\theta(x)$ and $\psi(x)$ relative to the period or to the distance between neighbouring defects $L_{pdw}$ decreases with approaching $T_{pdw}$. This means that the near $T_{pdw}$ the CPDW state consists of the almost isolated defects with the growing distance between them $L_{pdw}(T\to T_{pdw})\to \infty$.

 The probe function approach shows that the formation of CPDW state is determined by the competition between DMI and the effective  anisotropy energy including the  corrections in Eq.\ref{Eq:EnergyS} generated by the spin supercurrents. The anisotropy favours the homogeneous state while DMI favours the CPDW one.  To study this competition in detail we calculate the phase diagram as function of $T, \tilde{\alpha}_{SO}$ by comparing the energy of CPDW state with large $L_{pdw}= 20 \xi_0$ with that of the homogeneous states $F_{x,z}$. The result  shown in Fig.\ref{Fig:PDWphase}b demonstrates the
 regions where the ground state is CPDW and the homogeneous one. The solid black line shows the points of the transition from $|m_x|=1$ to $|m_z|=1$ homogeneous states where the effective anisotropy vanishes $F_x=F_z$.  Near this transition the effective anisotropy vanishes, so that the CPDW region is elongated along the black curve. 
 Due to this peculiar form  there appears an interval of $\tilde\alpha_{SO}$ where CPDW state has a re-entrance behaviour, e.g. with the lowering $T$ along the black dashed line in Fig.\ref{Fig:PDWphase}b. For other values of $\tilde \alpha_{SO}$ one can have a single transition into CPDW, e.g. along the red dashed lines.


 To summarize, we have discovered the CPDW hybrid states  combining the array of chiral domain walls and the modulation of the superconducting order parameter.   
The CPDW state is generated by the DMI mediated by Cooper pairs which can be considered as the contribution to the energy from superconducting spin currents. 
 The CPDW state have several distinct features as compared to the known non-homogeneous ground states.   
 {\bf (i)} The CPDW state breaks the parity symmetry having definite sign of magnetic chirality.  {\bf (ii)} The typical cryptoferromagnetic state has\cite{anderson1959spin, buzdin1988ferromagnetic, bulaevskii1985coexistence} the period  $\ll  \xi_0 $
 while for the CPDW it is  $ L_{pdw} \geq  \xi_0$ where $\xi_0$ is the Cooper pair size.    {\bf (iii)} The transition to the CPDW state occurs via the penetration of defects - chiral domain walls in contrast to the second order phase transition to the cryptoferromagnetic phase\cite{bergeret2000nonhomogeneous}.
  {\bf (iv)} 
  In contrast to the cryptoferromagnetic state which has been observed only in the exotic compounds\cite{bulaevskii1985coexistence}
such hybrid chiral states can be obtained in the existing FI/S structures such as YIG/Nb/Pt or EuS/Al/Pt multilayers.
   The CPDW state is also qualitatively different from the previously known  Fulde-Ferrel-Larkin-Ovchinnikov state\cite{fulde1964superconductivity,larkin1965inhomogeneous}and the phase-modulated superconducting state\cite{samokhin2005paramagnetic,kaur2005helical,Buzdin2008}  because in the CPDW state the proliferation of the inhomogeneity occurs coherently in the superconducting and magnetic subsystems.

  {\it Acknowledgements}
This work was supported by the Academy of Finland (Project No. 297439) and Russian science foundation. MAS acknowledges useful discussions with Tero Heikkilä and Yao Lu. The work of I.V.B has been carried out within the state task of ISSP RAS with the support by RFBR grant 19-02-00466. I.V.B. and D.S.R. also acknowledge the financial support by Foundation for the Advancement of Theoretical Physics and Mathematics “BASIS”.

 \appendix
 
 
 \section{General expression for spin supercurrent}
 
  The general expression for the free energy correction due to the superconductivity 
 is \cite{virtanen2020quasiclassical}
 \begin{align} \label{Eq:FreeEnergyGen}
  & F_S =
  \frac{\nu\pi T}{8} \sum_{\omega }
  {\rm Tr}
  [ D (\check{\nabla} \hat g)^2
  -
  4(\omega + i \bm h \bm{\hat\sigma} )\hat{\tau}_3 \hat g
  - 
  2 \hat{\Delta}\hat{g}  ]
 \end{align}
where $\check{\nabla}_k = \nabla_k - 
i [\cdot, {\cal A}^\gamma_{k}\hat \sigma_\gamma]/2$. 
 The  spin supercurrent in diffuse superconductor is given by
  \begin{align} 
 \label{Eq:SpinCurrentDiff}
 {\bm J}^\gamma =  \frac{i\pi \nu D}{16}
 T \sum_\omega
 {\rm Tr} ( \hat\sigma_\gamma \check{g} 
 \check{\nabla} \check{g} ) ,
 \end{align}
The direct calculation of the free energy (\ref{Eq:FreeEnergyGen}) variation over the SU(2) field ${\cal A}^\gamma_{k}$ shows the relation  Eq.~(\ref{Eq:SpinCurrentVar}) is valid in the general case.

 \section{Ginzburg-Landau free energy expansion}
 
 To derive the free energy of Pt/S/FI system we use 
 {\bf Usadel equation} in the presence of SOC assuming that the spin-splitting of electronic bands is much smaller than the Fermi energy. It is formulated in therms if the momentum-averaged Green function (GF) $\hat g$ which is the matrix in Nambu and spin space
 \begin{equation} \label{Eq:Usadel}
 \tilde{\nabla} (D \check g \tilde{\nabla} 
 \check g)=
 [\Delta\hat\tau_1 + \hat\tau_3( \omega + i {\hat{\bm\sigma}}\bm{h} ),  \check g],
 \end{equation}
 where 
 $\tilde{\nabla}_k\cdot=\partial_k\cdot-i[\cdot, \hat {\cal A}_k]/2$ is the SU(2)-covariant gradient and $D$ is the diffusion coefficient which can be layer-dependent.
 The spin-dependent gauge field $\hat {\cal A}_k = A_{ki}\hat\sigma_i$ exists in the normal metal region, the gap $\Delta$ and exchange field 
 ${\bm h}$ are
 non-zero in the superconductor and ferromagnet, respectively.

Close to the critical temperature we have the expansion 
(we consider $\omega>0$)
\begin{align}
\hat g= \hat\tau_3 + \delta \hat g
\end{align}
 where  
 $\delta \hat g$ is the expansion by 
 $\Delta$, $\nabla\Delta$ and ${\cal D}_k \bm m$. 
 To begin with, we find the first-order correction by $\Delta$  \begin{align}
  &  \hat g_\Delta = 
 \Delta   \hat\tau_1
  (\omega + i \bm h\bm \sigma )^{-1} 
 - 
 \frac{\Delta^3}{2}  \hat\tau_1
  (\omega + i \bm h\bm \sigma )^{-3} 
 - 
 \\ \nonumber
 & \frac{\Delta^2}{2} \hat\tau_3 
  (\omega + i \bm h\bm \sigma )^{-2}
 \end{align}
 We are interested only in the spin-singlet part anomalous part of GF which is 
  \begin{align}
  &   g_\Delta = 
 \Delta  \frac{\omega}{\omega^2 + h^2}
  - 
  \frac{\Delta^3}{2} 
   \frac{\omega^3 - 2 \omega h^2 }{(\omega^2 + h^2)^3}
   \end{align}
 
 In case if $|h|=const$ the gradient term in Eq.\ref{Eq:Usadel} yields corrections to the spin-singlet part of the GF
  \begin{align}
 &  g_{\nabla} =  
 \frac{(\omega^2-h^2) \nabla(D \nabla \Delta)}{(\omega^2 + h^2)^2} 
 - 
 \\ \nonumber 
 & \frac{   D
    \nabla_k  \Delta ({\cal{D}}_k \bm h) \bm h
    + 
    \Delta  {\cal{D}}_k( D {\cal{D}}_k \bm h ) \bm h
    }{(\omega^2 + h^2)^2}
    \end{align}
 
 In case if the absolute value of the exchange field is constant $|h|=const$ the above expression can be simplified 
 by taking into account that $ ({\cal{D}}_k \bm h) \bm h =0$, so that ${\cal{D}}_k (\bm h {\cal{D}}_k \bm h) ={\cal{D}}_k\bm h {\cal{D}}_k \bm h + \bm h {\cal{D}}_k {\cal{D}}_k \bm h=0 $. Then we get
 \begin{align}
 &  g_{\nabla} =  
 \frac{(\omega^2-h^2) \nabla(D \nabla \Delta)}{(\omega^2 + h^2)^2} 
 - 
  \frac{  D
    \Delta  ({\cal{D}}_k \bm h )^2
    }{(\omega^2 + h^2)^2}
    \end{align}

 The self-consistency equation reads 
 \begin{align}
     \tau \Delta = - \pi T \sum_{\omega>0} ( g_\Delta + g_\nabla  - \Delta/\omega)
 \end{align}
 where $\tau = 1- T/T_c$. 
 Introducing the dimensionless order parameter $\psi=\Delta/T_c$ we get the GL equation  
 \begin{align}
 &  a \psi - a_m ({\cal{D}}_k \bm m)^2 \psi  
    + b \psi^3 
    -  \nabla (c \nabla \psi) = 0 
     \\
    & a/\nu T_c^2 =    \pi T_c \sum_{\omega>0} \frac{h^2}{\omega (h^2 + \omega^2) }
    - \tau  
    \\
    & b/\nu T_c^2 = \pi T_c^3 \sum_{\omega>0} \frac{\omega^3-3\omega h^2}{2 (h^2 + \omega^2)^3}
    \\
    & 
     c/\nu T_c^2\xi_0^2 =  \pi T_c^2 \sum_{\omega>0} \frac{\omega^2-h^2}{ (h^2 + \omega^2)^2}
    \\
    & a_m/\nu T_c^2 \xi_0^2=   \pi T_c^2 \sum_{\omega>0} 
    \frac{ h^2}{ (h^2 + \omega^2)^2}
 \end{align}
 At $h=0.5 T_c$ the numerical values are 
 $\tilde a =  0.026-\tau$, $\tilde b = 0.046$, 
$\tilde c =0.37$, $\tilde a_m = 0.008$.
 
 Since the GL equation is obtained from the condition 
 $\delta F_S/\delta \psi =0$ 
 we get the expression for the free energy GL functional 
 \begin{align} 
     F_S  =  a \psi^2 - \frac{a_m}{2} ({\cal{D}}_k \bm m)^2 \psi^2  +
     \frac{b}{2} \psi^4  + c D (\nabla \psi) ^2
 \end{align}

 \section{Proof of the CPDW state existence with the help of the probe function}
 
 To demonstrate the possibility of the chiral state formation we consider the probe function with the homogeneously rotating magnetization 
 $m_x=-\sin(qx)$, $m_y=\cos(qx)$ and $\psi= const$.
 Substituting this probe function into the Eqs.7,8 in the main text and averaging this by the period of modulation $2L_{pdw} =2\pi/q$ we get
 \begin{align} \label{AppEq:FCPDW}
 & F_{CPDW} =  q^2 + \frac{1}{2} +
     \\ \nonumber
 & \left [\tilde a-\frac{\tilde a_m}{4}(2q^2 + 4 \tilde\alpha_{SO} q  + \tilde\alpha_{SO}^2) \right]\psi^2  + \frac{\tilde b}{2} \psi^4  \end{align}
 This energy is to be compared with the energies of the two possible homogeneous stated 
 \begin{align} \label{AppEq:Fx}
  &   F_{x} =   \tilde a \psi^2  +
     \frac{\tilde b}{2} \psi^4  
     \\  \label{AppEq:Fz}
  &   F_z =  1  + 
  ( a -\tilde \alpha_{SO}^2 \tilde a_m/2) \psi^2  +  
  \frac{\tilde b}{2} \psi^4
 \end{align}

 After minimizing Eqs.(\ref{AppEq:FCPDW}, \ref{AppEq:Fx}, \ref{AppEq:Fz}) by $\psi$ we get 
 \begin{align}
  &   F_{CPDW} = 
     \\ \nonumber
  &   \frac{1}{2}  - \frac{1}{2\tilde b}\left [a-\frac{\tilde \tilde a_m}{4}(2q^2 + 4 \tilde\alpha_{SO} q  + \tilde\alpha_{SO}^2) \right]^2 +   q^2 
  \\
  & F_{x} =   -\frac{\tilde a^2}{2\tilde b} 
   \\
  &   F_z =  1  -  
  \frac{( \tilde a - \tilde\alpha_{SO}^2 \tilde a_m/2)^2}{2\tilde b}
 \end{align}
First, one can see that the transition from $\bm m =\pm  \bm x$  to $\bm m =\pm  \bm z$ 
state occurs at the temperature defined by  $F_z<F_x$ which yields 
 \begin{align}
 \tilde a < \frac{\tilde a_m\tilde \alpha_{SO}^2}{4} - \frac{2\tilde b}{\tilde a_m \tilde \alpha_{SO}^2}
 \end{align} 
 The CPDW exist if $F_{CPDW}(q=0) < min(F_z,F_x)$
which yields the condition defining the temperature interval 
 \begin{align}
 \frac{3 \tilde a_m \tilde \alpha^2_{SO}}{8} - \frac{3\tilde b}{ \tilde a_m\tilde \alpha^2_{SO}}  < \tilde a < 
 \frac{\tilde  a_m\tilde \alpha^2_{SO}}{8} - \frac{2\tilde  b}{ \tilde a_m\tilde \alpha^2_{SO}} 
 \end{align}
 
 Thus, one can see that with lowering temperature and thus decreasing $a$ we first encounter the homogeneous magnetization rotation transition to
 $\bm m = \pm \bm z$ state and after that there is a transition to CPDW.

  \bibliography{refs2}

 \end{document}